# Network-thinking to optimize surveillance and control of crop parasites. A review

Andrea Radici[1*], Daniele Bevacqua[2], Leonardo Miele[2], Davide Martinetti[1]

**Affiliation** [1]INRAE, UR 546 BioSP, Site Agroparc, 84914 Avignon, France. [2]INRAE, UR 1115 PSH, Site Agroparc, 84914 Avignon, France.

## Abstract

Increasing cultivated lands, crop homogenization and global food trade have fostered the spread of crop pests and diseases. Optimizing crop protection is urgently needed to ensure food safety. One aspect of crop protection is surveillance, which focuses on the early detection of a parasite, and control, aiming to fight and possibly eradicate it. Network theory has been widely used to model the spread of human and animal infectious diseases in systems described through nodes and edges. It has been successfully used to optimize monitoring and immunization campaigns. In crop protection, there is a growing literature using this theory to describe parasites spread and to conceive protection strategies.

Here we review the use of network theory in crop protection, from the more descriptive to the more applied approaches aimed to optimize crop protection. We retrace the logical process that has led epidemiological models to rely on network theory, and we provide examples of how the spread of crop parasites has been represented via a network description. We define the objectives of surveillance and control, and we show how these have been declined in the network-based epidemiological sphere and then adapted in the agricultural context. We eventually discuss the discrepancy between the application of network theory in surveillance and control to identify culprits and solutions.

We find that: *i*) scientists have successfully interpreted very different modes of parasitic transmission under the lens of network theory; *ii*) while network-based surveillance has progressively clarified its objectives and sound tools have been proposed, network-based control has been less studied and applied; *iii*) network-thinking must address how to properly define edges and nodes at different geographic scale to broad its application in crop protection.

**Keywords**: crop protection, networks, parasites, epidemic surveillance, epidemic control.

---

[*] Contact: andrea.radici@inrae.fr

# Contents

## 1   Introduction

Pests and pathogens are responsible for a reduction between 17% and 30% of crop production at the global scale (Savary et al. 2019). In the next decades, crop losses are expected to increase due to the narrowing of diversity in crop species and the increase of food demand, pesticide resistance, and of global trade which determine favorable conditions for pests and diseases spread as they increase the host abundance and contact probability between hosts (Carvajal-Yepes et al. 2019; Khoury et al. 2014; Ristaino et al. 2021a). Climate change could also facilitate the spread of certain pathogens and pests (Corredor-Moreno and Saunders 2020) and most of the losses are expected in countries with expanding populations, where food supply is already an issue (Adam 2021; Tilman et al. 2011). Moreover, agriculture is asked to mitigate its environmental and health impacts related to the use of chemicals which have been the major tool to protect crop from pest and disease in the last decades (Tudi et al. 2021; Vitousek 1994). The challenge of reconciling sufficient food production with social and environmental viability is acknowledged as part of the Sustainability Development Goals of the UN 2030 Agenda (Foley et al. 2011; Lee et al. 2016; Tilman et al. 2002).

Therefore, surveillance and control, which are complementary measures to protect crops from parasites, hereinafter intended for both pests and pathogens, ought to be optimized (Carvajal-Yepes et al. 2019; Morris et al. 2022; Ristaino et al. 2021). Surveillance, which is mostly based on data collection and analysis, aims to detect the presence of parasites early and to know the phytosanitary status of a crop. Control aims to reduce parasitic dispersal, either by immunizing hosts, *e.g.* by cultivating resistant varieties, or by limiting parasites dispersal by reducing contacts between individuals. An effective crop protection strategy requires both surveillance and control to work together: in fact, the efficacy of control measures critically depends on the timing of detection of the parasite, which is likely to spread if favorable conditions are present.

Network theory, originally intended to study properties of systems described by a set of nodes connected via edges (Newman 2003), has been successfully applied to epidemiology. In this context, the nodes of the network represent host individuals or distinct host populations and the edges represent connections between nodes. In recent decades, network theory has been extensively used to model the spread of infectious diseases in humans and animals and to inform protection policies (Brockmann and Helbing 2013; Dubé et al. 2011; Kao et al. 2007; Keeling and Rohani 2011), the recent Covid-19 pandemic being a paradigmatic case (Block et al. 2020; Della Rossa et al. 2020; Saunders and Schwartz 2021).

Network theory has proven to be a valid tool in crop protection to model disease spread and to design surveillance and control strategies (Cunniffe et al. 2015; Garrett et al. 2018; Moslonka-Lefebvre et al. 2011; Parnell et al. 2017). Unlike animals and humans, plants do not move, so that the transmission of a parasite depends on its displacement. This can be active, such as flying from an orchard to another, or passive, driven by living agents (called vectors) or abiotic elements such as air masses, water currents, or even via transport of goods. In the case of crop parasites, nodes of the network usually identify locations where the host is present, which generally correspond to cultivated areas (Fig. 1).

An example of network-thinking in crop protection is the identification of those locations that contribute the most to spreading parasites (Zhang et al. 2016). This identification, hereinafter called prioritization, is needed because crop protection is expensive, so it is important to rank and select a reduced set of locations (nodes) to be inspected for surveillance and/or to be treated for control (as in Fig. 1d). Recently, more advanced research has been conducted to investigate useful node properties which allow this prioritization (Hernandez Nopsa et al. 2015; Martinetti and Soubeyrand

2019; Sutrave et al. 2012) in the case of re-emerging parasites, such as *Puccinia graminis* (Meyer et al. 2017) or *Xylella fastidiosa* (Strona et al. 2017) in Europe. The implications of networks for crop protection are rising (Garrett et al. 2018; Jeger et al. 2007; Moslonka-Lefebvre et al. 2011; Parnell et al. 2017; Shaw and Pautasso 2014), but little research has focused on rigorously clarifying their properties for surveillance and control (Holme 2017). Also, the level of application of network-thinking is not the same in surveillance and control: this review aims to present the state of the art in terms of network-thinking in crop protection, investigate the main obstacles to applications and delineate research perspectives.

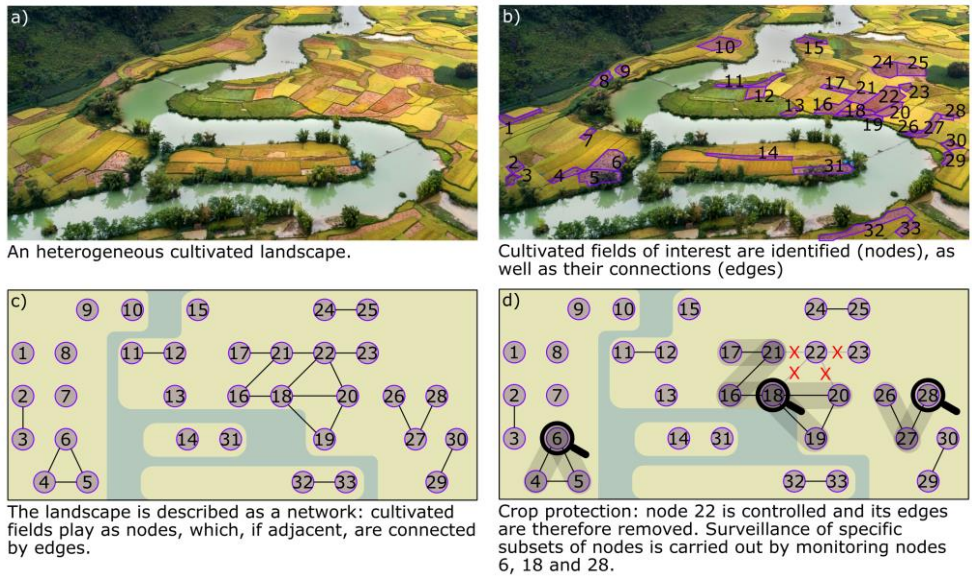

***Fig. 1*** *– A network of scattered fields over which crop protection is conducted. a) Fields of interest are part of a complex landscape; b) in this example, we imagine that the reddish fields play as nodes, and that that parasites can move from a field to an adjacent one (via an edge); c) the network is formalized into a graph made of a set of nodes connected via edges, discarding all the trivial elements (only the river is represented in the background to help readability); d) control of node 22 (via cultivation of resistant cultivars and/or phytosanitary treatment) prevents parasitic flow in and throw this node: its edges are therefore removed. The presence of parasites in the network is acknowledged via surveillance of nodes 6, 18 and 28 (this latter, for example, allows to indirectly monitor connected nodes 26 and 27).*

## 2  Networks: definitions and properties

Networks are sets of nodes connected via edges (Börner et al. 2007). In epidemiology, nodes represent hosts (individuals or populations), while edges represent infectious contacts. A classic way to represent a network is a square matrix **W**, called adjacency matrix, whose generic element $w_{i,j}$ represents the edge connecting node $i$ to node $j$. Edges can be either binary ($w_{i,j} = 1 \lor 0$) or associated to weights, generally positive ($w_{i,j} \in \mathbb{R}^+$); either undirected, if a relation from a node $i$ to a node $j$ implies the same from $j$ to $i$ (it follows that the adjacency matrix is symmetric: $w_{i,j} = w_{j,i}$) or directed; lastly, they can be dynamic (edges arrangement and weight vary with time) or static.

Nodes can be characterized in terms of their features, which are quantified via metrics (often called centralities). The simplest metric that one can build is the degree, usually indicated by $k$, *i.e.* the number of edges connected to a node. In the case of weighted networks, the notion of degree is usually replaced by that of strength, *i.e.* the sum of the weights of the edges connected to a node. In the case of directed networks, it can be further distinguished between in- and out-degree/strength, considering only in- or out-going edges. The probability distribution of $k$ in a given network is called degree distribution, and its average is indicated as $\bar{k}$. The degree distribution can be used to classify network properties as a whole. For instance, homogeneous networks are built in such a way that all nodes have the same degree $\bar{k}$. On the other hand, the Barabási-Albert algorithm allows to build scale-free networks (Barabási and Albert 1999), in which the degree distribution follows a power law.

Several node metrics can be defined (Mastin et al. 2020), such as the betweenness of a node $w$, i.e., the number of shortest path connecting whatever couple of nodes of the networks $i$ and $j$ and passing through $w$ (Barabási 2016). Interested readers could find an exhaustive summary of node metrics in Lü et al. (2016).

## 3  Modelling epidemics

Mathematical epidemiological modelling has its origin in the S-I compartmental model (Kermack et al. 1927). In its simplest formulation, this assumes that a population is divided into susceptible (S) and infected (I) individuals and describes its temporal variation via a system of differential equations:

$$\begin{cases} \dot{S} = -\beta SI + \mu I \\ \dot{I} = \beta SI - \mu I \end{cases} (1)$$

Where $\beta$ represents the disease transmission rate, that drives S to become I, and $\mu$ represents the recovery rate, that drives I to return to S. This model can be extended to include other processes (*e.g.* immunity, natural mortality and fertility) and it allows to define the basic reproductive number $R_0$, which indicates "the expected number of secondary cases which one case would produce in a completely susceptible population" (Dietz 1993). In the model described by Eq. 2, $R_0$ is:

$$R_0 = \frac{\beta}{\mu} (2)$$

If $R_0$ of a disease is greater than 1, the introduction of an infected individual into a fully susceptible population will cause the spread of the disease, otherwise it will progressively die out. This model assumes homogeneous mixing, which means that individuals have equal probabilities of transmitting the disease to everybody else. This is not the case, for example, when the spatial location of individuals defines their contact structure, which can be formally represented by a network (Pastor-Satorras et al. 2015). In that case, classic results of epidemic theory may not apply (Pellis et al. 2015).

Moreover, networks can embed SI dynamics. In a time-discrete framework, a node can be either susceptible or infected; at each time step $\delta$, each infected node $i$ can infect its $k$ neighbors at a probability $\beta\delta$, and recover with a probability $\mu\delta$. For instance, in SI

dynamics on homogeneous networks it can be shown (Boccaletti et al. 2006) that Eq. 2 is modified into:

$$R_0 = \frac{\bar{k}\beta}{\mu} \quad (3)$$

So $R_0$ depends explicitly on the network via the average degree $\bar{k}$. For non-homogeneous networks, $R_0$ can be recomputed (Boccaletti et al. 2006) instead as:

$$R_0 = \frac{\overline{k^2}\beta}{\bar{k}\mu} \quad (4)$$

Where $\overline{k^2}$ is the 2nd moment of the degree distribution of the network. It has been shown that, for epidemic processes in scale-free networks, $\overline{k^2}$ tends to infinite (Pastor-Satorras and Vespignani 2001): the disease will spread due to the network topology provided a transmission rate $\beta > 0$ (Jeger et al. 2007; Shaw and Pautasso 2014). Eq.s 3 and 4 suggest that network-related control strategies can help decreasing $R_0$ below 1 by acting on the network structure.

## 4 Crop-parasite interactions in a network framework

In crop epidemiology, plants are connected via parasitic dispersal (Shaw and Pautasso 2014). Parasites can move over extremely long distances through different means (air, water, soil, insects; Aylor 2017; Jordano 2017) or can be assisted by human activities (Garrett et al. 2018; Harwood et al. 2009; Hulme 2009; Santini et al. 2018). The definition of the network depends on the scale and mechanism of the contacts (Garrett et al. 2018; Gilligan 2008; Shaw and Pautasso 2014). Below is a list of examples on how networks have been used to represent and study crop-parasite interactions.

### 4.1 Vector-transmitted parasites

Strona et al. (2017) studied the spread of the pathogenic bacteria *X. fastidiosa* in an olive-orchard-dominated landscape, via its main European vector, the meadow spittlebug *Philaenus spumarius* (Martelli et al. 2016). They considered olive orchards as nodes, which are connected by an edge if they are within 1 km "between the two closest sides", since observations suggested that this is the order of magnitude of adult spittlebug daily flights. They obtained a binary network, since edges are not associated to a weight, and undirected, since connection among two orchards are always reciprocal. Instead, de la Fuente et al. (2018) built a more complex network by modelling the spread of the alien nematode *Bursaphelenchus xylophilus*, causal agent of pine wilt disease, via the longhorn beetle *Monochamus galloprovincialis*. They considered coniferous forests as nodes connected via weighted, directed and time-varying edges. Each edge represents the yearly probability of transmission of the disease from node $i$ to node $j$, whose weight decreases with the distance between $i$ and $j$ and increases with the assumed number of nematodes in $i$ in the previous year.

### 4.2 Wind-dispersed

Sutrave et al. (2012) studied a network of soybean fields in the US, susceptible to the airborne pathogen *Phakopsora pachyrhizi*, obtaining a similar network to that of de la Fuente et al. (2018). In this model, each node represents an administrative county, while each edge's intensity is determined by the daily average wind intensity and direction, increases with the density of soybean in the connected nodes, and decreases with the distance. In this case, the network is weighted, directed (because of the prevailing wind direction) and dynamic. Enlarging the scale to the continental one, Meyer et al. (2017) used Lagrangian simulation of air masses to model the airborne dispersal of the fungal pathogen *P. graminis*, causal agent of stem rust of wheat, among African and Asian countries. Countries play as nodes, while the daily mean proportion of airborne spores successfully transmitted from a country to another define the weighted and directed edges.

### 4.3 Human-mediated

A recent study by Andersen et al. (2019) models the spread dynamics of a hypothetical pathogen affecting sweet potato on a network accounting for local seed trade in Uganda. In such networks, nodes are provided by economic actors or their grouping (sellers, villages), whose interaction, in this case "at least one transaction during the growing season", represent directed edges. Hernandez Nopsa et al. (2015) used the railway networks in the US (and Australia) to model the potential spread of arthropods between grain elevators used to store wheat. In their framework, sets of elevators located in the same state play as nodes, while directed and weighted edges are proportional to the volume of wheat moved from a state to another via railway, which traditionally account for more than 70% of the shipped wheat in the US.

### 4.4 Multi-mechanism

Eventually, networks may also summarize multiple mechanisms involved in parasitic spread (Garrett et al. 2018). For instance, nodes may either be farms, where crops are produced, or warehouses, where crops are stored after harvesting (Fig. 2). Edges, representing possible pathways of parasitic spread, can be subdivided into retail sub-network, representing purchase, and landscape sub-network, representing physical movement by wind (or insects or animals). Retail sub-network may be undirected, working both ways between producers and retailers via returnable shipping crates, while landscape sub-network may have a fixed direction, because of prevalent wind.

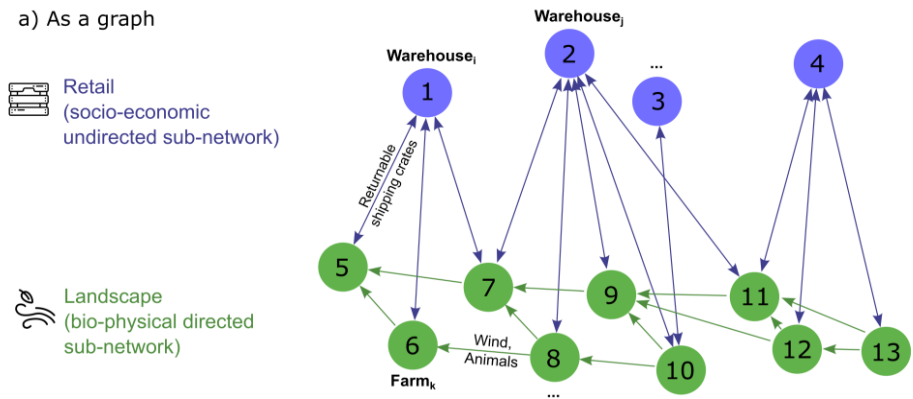

*Fig. 2* – A multi-mechanism epidemic network. Nodes are either farms, where crop is produced, or warehouses, where harvested crop is stoked. Edges, representing possible parasite spread pathways, can be subdivided into two sub-networks: retail, representing purchase, and landscape, representing physical movement by wind insects, or animals. Retail sub-network is undirected, since it may work both way between producers and retailers (for example, via returnable shipping crates), while landscape sub-network has a fixed direction, (for example, because of prevalent wind). Network is represented a) as a graph, b) as an adjacency matrix, in which the retail sub-network is symmetric (in blue) while the landscape one is asymmetric (in green).

## 5 Surveillance

Surveillance can be defined as the "collection and analysis of information for detection and successful control of emerging pathogens" (Parnell et al. 2017). Since this collection is expensive, one needs to prioritize a reduced set of locations for monitoring to find clues of crop disease emergence.

## 5.1 Prioritize nodes to be monitored

In a network-based surveillance, prioritization of nodes to be monitored (also referred as sentinels) depends on the topology of the network structure. Surveillance may be divided into different objectives (Herrera et al. 2016; Holme 2018). For surveillance of emerging diseases, Parnell et al. (2017) distinguished between *i*) *detection*: understanding weather a pathogen has arrived in an area, with the minimum possible delay; *ii*) *estimation*: estimating the proportion of diseased individuals (incidence) at a given time; *iii*) *targeting*: maximizing the detection of new cases. These objectives are not pursued without trade-offs. For example, in the case of an already detected parasite, increased "targeting" would imply monitoring those nodes which are connected to the infected one, to the detriment of "estimation". In contrast, good "estimation" can be obtained via homogeneous monitoring, *e.g.* via random sampling of the entire network, but would result in a worse "targeting".

Different prioritization algorithms allow for the pursuit of different objectives. Herrera et al. (2016) compared performances of different algorithms to determine a sentinels set in epidemic dynamics on social networks. The criteria for prioritization are: "most connected" (nodes characterized by the highest degree); "random"; "random acquaintance" of random individuals, *i.e.* a random node connected to a randomly extracted node. As expected, sentinels identified via the "random" strategy yielded indications representative of the overall populations, with early warning close to zero (there was no time lag between the surveillance subset and the entire population to reach 1% prevalence) and peak ratios close to one (i.e., peak in incidence was almost identical in the surveillance subset and the entire population). Degree-based strategies, such as "most connected", consistently provided the earliest warnings. Herrera et al. explored different network topologies and observed that surveillance performances tend to be highly differentiate the less homogeneous the network is.

## 5.2 Techniques for crop surveillance

Several studies have shown that traditional site prioritization techniques for crop surveillance, based on historical recordings of past epidemics or more complex approaches (Parnell et al. 2017), may be integrated (Sanatkar et al. 2015; Sutrave et al. 2012) or even outperformed (Martinetti and Soubeyrand 2019) by network metrics.

Sutrave et al. (2012) explicitly explored network metrics for optimizing surveillance in a continental dynamic network among soybean producing counties. The authors explored several sentinel prioritization strategies. They observed that surveillance prediction errors were minimized by combining strength and "infection-frequency" (a metric considering past presence of the disease). In a recent study, Radici et al. (2022) simulated worldwide transport of airborne spores of another cereal pathogen, *P. graminis*, among squared wheat-cultivated cells ($\approx 2,000\ km^2$) connected by air-mass trajectories computed via a Lagrangian model, obtaining a dynamic directed and weighted connectivity network. They showed that a surveillance strategy based either on betweenness or Set cover algorithm (a modified version of in-degree) minimizes the detection delay of simulated outbreaks. Furthermore, they showed that, given the high density of the epidemic network, surveillance strategies based on separate subset of nodes (representing countries) lead to a sub-optimal allocation of sentinels with respect to considering the complete set of nodes.

Network-based prioritization has been also applied in the case human-mediated crop parasites. Andersen et al. (2019), who studied the spread dynamics of a hypothetical pathogen affecting sweet potato, suggested that degree and betweenness may behave as well as other more complex metrics in node prioritization. Their importance was already noted in previous research on arthropods detection in grain networks (Hernandez Nopsa et al. 2015). Similarly, Buddenhagen et al. (2017) considered a local multi-mechanism potato trade network in Ecuador, and stated that high in-and out-degree nodes were to be surveilled first.

Recent research has been conducted to build more complex network metrics and assess their suitability for surveillance compared to traditional ones. For instance, Martinetti and Soubeyrand, (2019) followed an approach similar to that of Herrera et al. (2016) to compare several surveillance options in the case of *X. fastidiosa* in southern France. In this case, they divided the region into 1 $km^2$ cells, which represent nodes of the network, while edges' weights are built on notion of risk: each weight $w_{i,j}$ is the product of a previously computed risk indicator $r_i$, $r_j$ in the connected nodes $i$, $j$. They tested prioritisation techniques based on network metrics (such as k-shell, Kitsak et al. 2010; VoteRank, Zhang et al. 2016; generalized random walk accessibility, or GRWA, Herrera et al. 2016) and other metrics, such as risk-based and random, to maximize the "detection" (*sensu* Herrera et al. 2016). Strategies based on VoteRank, GRWA and risk-based furnished the earliest detections.

# 6 Control

Control can be defined as the set of measures aimed to minimize the disease size or the occurrence of losses within a diseased population (Keeling and Rohani 2011). Network based epidemic control consists in the preventive removal of a set of nodes (or of their edges) from the epidemic network, in the hope that this will minimize the spread (Shaw and Pautasso 2014).

## 6.1 Prioritize nodes to be removed

Traditional disease control methods at the population scale, including vaccines(Keeling and Rohani 2011), may rely on network knowledge. Vaccination levels for reducing transmission usually depend on a threshold determining herd immunity (Fine 1993), and are in turn calculated via $R_0$. This approach allows a first estimation of the immunization level needed to stem the epidemics in homogeneous networks. As effect of the vaccination of a proportion $g$ of the nodes, $R_0$ can be tuned below 1, thus extinguishing the spread of the disease (Eq. 5):

$$R_0 = \frac{\bar{k}\beta(1-g)}{\mu} \quad (5)$$

Network-based vaccination strategies have been studied to stem epidemics on both human and animal populations (Rushmore et al. 2014). It has been shown that neglecting highly heterogeneous networks structures may lead to ineffective or inefficient immunization strategies (Jeger et al. 2007). Identification of the nodes to be immunized via simple metrics, such as node strength, proved to help reducing the vaccine coverage threshold, as in the case of wild chimpanzees (Rushmore et al. 2014). Alternatively to vaccination, isolation (Keeling and Rohani 2011) may as well target specific nodes (*i.e.* isolation of an infected node by removing all its edges) or random (*i.e.* removal of a fraction of edges independently on the nodes health status, such as social distancing in the case of Covid-19). The first case corresponds to increasing $g$, equivalently to an immunization or vaccination. In the second case, the strategy affects $\bar{k}$, whose reduction can bring the value of $R_0$ below 1 (Block et al. 2020).

Several authors highlighted that control strategies should concentrate in immunizing highly connected nodes (Jeger et al. 2007; Lloyd-Smith et al. 2005; Pastor-Satorras and Vespignani 2002; Shaw and Pautasso 2014). In the case of human diseases, targeted immunization proved to achieve good results with rubella and mumps, whose heterogeneous transmission structure possesses scale-free properties (Pastor-Satorras and Vespignani 2001), rather than with pertussis, whose transmission structure is more homogeneous (Trottier and Philippe 2005). A comprehensive literature exists to support targeted action against sexually transmitted diseases, based on the scale-free properties of sexual partnership networks (Jeger et al. 2007).

Prioritization methods for targeted removal have been examined according to various network metrics, from the simplest (degree, betweenness or closeness, *i.e.* the reciprocal of the average distance from that node to all the other nodes; Freeman 1978; Sabidussi 1966) to more complex (random-walker based methods, Zhang et al. 2016). Methods based on k-shell decomposition have proved to produce optimal results compared to classic metrics in identifying influential spreaders (De Arruda et al. 2014; Kitsak et al. 2010), and some generalizations to dynamic networks have been explored (Ciaperoni et al. 2020; Galimberti et al. 2018). In the specific case of spatial networks, where weights have the meaning of spatial distance, generalized accessibility metrics (Travençolo and Costa 2008) have proved to be particularly suitable for targeted removal (De Arruda et al. 2014).

Despite the growing research on network-based disease protection strategies, there is a lack of clarification on how to distinguish metrics for determining suitable nodes for surveillance from those suitable for control (Holme 2017), which can occasionally overlap. Many of the before-mentioned metrics to prioritize nodes to immunize for control are common to those to be detected for surveillance. However, in principle, nothing guarantees that an optimal set of sentinels coincides with an optimal set of candidate nodes to be immunized, and vice versa, since surveillance and control are linked to different properties. Fig. 3 shows an example of a SI model running on a network, in which in-degree is used to optimize sentinels' identification and betweenness is used to identify the best node to immunize.

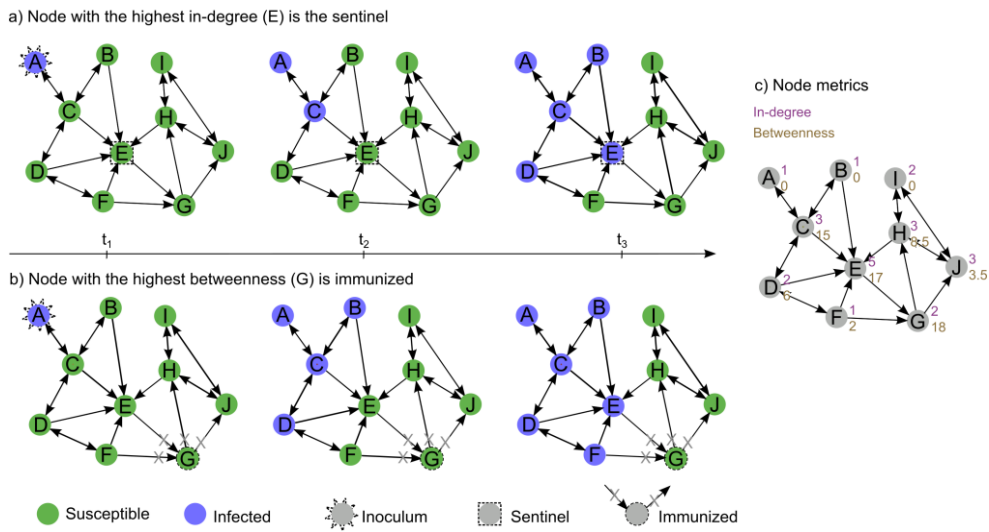

*Fig. 3* – *A network representing potential epidemic pathways and strategies based on node metrics to face parasites spread. In panels a) and b) the same SI epidemic process starting from node A (i.e., the inoculated node) is shown. At each time step, each susceptible node becomes infected if it has at least one infected neighbor pointing to it, without recovery. In panel a) node E is a sentinel: it easily detects the disease (i.e., it becomes infected) because of its high in-degree (i.e., the number of edges pointing to it). In panel b) node G is immunized through targeted removal of its edges: this way, any outbreak that starts from the left sub-network (nodes A, B, C, D, E, F) cannot reach the right sub-network (I, H, J), because of its high betweenness. Note that removal of node E result in poor control performances, since its removal is irrelevant with respect to disease spread. Similarly, G is an inefficient sentinel, since it has a low degree and it is reached late in an epidemic process. Panel c) shows the values of in-degree and betweenness computed for each node.*

## 6.2 Application to crop protection

Optimal crop parasites control methods, *e.g.* allocation of phytosanitary treatments, biological control, resistant varieties or even eradication of suspected hosts, only occasionally is explicitly investigated via network properties, as instead it is the case of targeted vaccination or isolation in humans and animals (Jeger et al. 2007). In one of these example, Strona et al. (2017) found that removing nodes with the highest PageRank (Page 1998) allows to reduce the size of the largest connected component (the largest set of nodes for which at least one path to any other node in the set exists) faster that using the degree or a random sampling, thus reducing the maximum spread of *X. fastidiosa*.

By contrast, some rules of thumb (*sensu* Chadès et al. 2011) loosely related to network-thinking have found greater application. For instance, it has been suggested that prevention planning may discourage growth of certain crop in marginal areas, which afford little profit but may provide chance for disease flow (Margosian et al. 2009). In this case, one could visualize main production areas as nodes, connected by edges representing marginal areas which allow the spread of the disease. Under this perspective, this disease control strategy translates into an edge-removal problem. Shaw and Pautasso (2014) reported successful experiences following this recommendation in reducing yellow rust of wheat, provoked by *P. striiformis*, in the northwest of China (Lu et al. 2011). It has been commonly observed that group of crops can remain uninfected if surrounded by immune individuals, as in the case of the soil-borne pathogen *Rhizoctonia solani* (Bailey et al. 2000). In this case, cultivated areas can play as nodes, immune areas as immunized nodes, while potential disease pathways as edges connecting them. In general, breaking the mono-cultural continuum into disconnected components reduces the opportunity of short-distance pathogen spread (Elton 1958). Given the definition of betweenness, we argue that this metric may support the identification of those nodes whose removal would more efficiently break the crop network (Fig. 3b).

These rules of thumb could gain more consistency if accompanied by network-based prioritization methods, such as a combination of different network metrics. Xing et al. (2020) evaluated the epidemic network given by a hypothetical pathogen affecting major crops across the world and suggested that an index weighting several network metrics (strength, nearest neighbors' degree, betweenness, eigenvector centrality) may give a comprehensive view of the nodes' contribution to connectivity. Additionally, Andersen et al. (2019) suggested that, as well as for surveillance, node characterized by high (out-)degree may be identified as influential spreaders and so prioritized to be immunized.

# 7   Conclusion

Although network-thinking is finding wider application in crop protection, its level of diffusion is unbalanced among surveillance and control. Network metrics, such as degree, strength, betweenness, VoteRank, and GRWA are more diffused to support surveillance, while control design in agriculture seems to require a multidisciplinary approach.

This imbalance may be due to several factors. A typical problem about the application of theoretical network tools in real cases is the adequately definition of nodes and edges. For example, network algorithms are very often conceived for unweighted, undirected and static networks, while real epidemic networks are frequently asymmetric, weighted and/or temporal. Generalizations may miss or distort the original meaning. To meet this need, a growing amount of research has been conducted to broaden the application of these metrics to real matrices (Fagiolo 2007; Galimberti et al. 2018; Holme 2005; Wang et al. 2008), declining them depending on the physical meaning of the edges, but generally with a loss of straightforwardness.

The challenge of spatial scale is an inveterate obstacle that network-thinking must address in order to expand its applications in agriculture, as pointed out by Shaw and Pautasso, (2014). Edges may be defined to describe extremely long-distance transport events along continents, in the order of $10^7$ $m$ (8,000 $km$ being distance from South Africa to Australia, perhaps the farthest documented periodical incursion of *P. graminis*; Visser et al. 2019) as well as cellular interaction between the hosts and its pathogens, on the order of $10^{-5}$ $m$ (26.4 $\mu m$ being the spore diameter of *P. graminis*; Eversmeyer and Kramer 2000), passing through the scale of a host ($10^0$ $m$ being size of a wheat stem) or a population ($10^3$ $m$ being the order of the size of an average US farm; Ritchie and Roser 2022). Moreover, the geographic scale of a node has crucial consequences for the meaning of applying network-based protection practices: to surveil/control a node means to completely monitor/immunize all the hosts within. Eventually, the choice of a node size is a compromise between the resolution of pathogen dispersal mechanisms (setting the biophysical constraints of the network) and the protection measures applicability (setting the management constraints of the network), in addition to ensure computational viability.

Broadening the scope of investigation, social acceptance may be another reason for the gap in application of network metrics among surveillance and control strategies. This gap may generally reflect a greater social impact of preventive crop control policies in a broad sense, not necessarily related to networks (Eriksson et al. 2019). Whereas optimal surveillance simply requires to geographically arrange the monitoring activities across a network, control thought node treatment implies an invasive intervention, such as the application of phytosanitary products or even the eradication of healthy hosts. Social acceptability of policies aimed at controlling alien pathogens invasions, as *X. fastidiosa* (Strona et al. 2017), relies on direct and indirect risk experiences and problem awareness of stakeholders, which may lead to a social underestimation of the probability of an invasion versus the concrete loss of a cultivated unit. Understanding the factors driving social acceptability of crop protection practices may be the key to the implementation of effective control methods (Marzano et al. 2017), in particular network-based strategies, with a positive feedback on the theoretical research, and *vice versa*.

## Acknowledgements

This research was supported by the Beyond project, funded by the French National Research Agency (ANR), the SuMCrop Sustainable Management of Crop Health Program and the Tom'Health (2021-2024) program of INRAE. We thank Cindy Morris for our fruitful discussions and Sabrina Volponi for valuable language tips. We thank ThuyHaBich, the author of the photo in Fig. 1a, which represents a Vietnamese countryside. We downloaded it freely from https://pixabay.com/.

## Author contributions

## Fundings

The authors acknowledge the support of funding from the French National Research Agency (ANR) for the BEYOND project (contract # 20-PCPA-0002), SuMCrop Sustainable Management of Crop Health Program and the Tom'Health (2021-2024) program of INRAE (which supports the work of Leonardo Miele).

# Declarations

**Conflict of interest.**

The authors declare no competing interests.

**Ethics approval**

Not applicable.

**Consent to participate**

Not applicable.

**Consent for publication**

Not applicable.

**Data availability**

Not applicable.

**Code availability**

Not applicable.